\documentclass{jetpl}
\usepackage{cite}
\usepackage{xcolor}
\twocolumn

\lat

\title{Complex dynamics of optical solitons interacting with nanoparticles}

\rtitle{Complex dynamics of optical solitons interacting with nanoparticles}

\sodtitle{Complex dynamics of optical solitons interacting with nanoparticles}

\author{D.\,A.\,Dolinina, A.\,S.\,Shalin, A.\,V.\,Yulin\thanks{e-mail: alex.v.yulin@gmail.com}}

\rauthor{D.\,A.\,Dolinina D.A., A.\,S.\,Shalin A.S., A.\,V.\,Yulin A.V.}

\sodauthor{ Dolinina, Shalin, Yulin,}

\address{ITMO University,
197101 Saint-Petersburg, Russia}

\dates{}{}

\abstract{In this Letter we further develop the proposed approach of optical manipulation based on the interactions of non-linear optical effects with nanoparticles. The interaction of the dissipative optical solitons with nanoparticles is studied numerically. It is shown that the attraction of the nanoparticles to the solitons can result in the formation of a stable bound state of the solitons and the nanoparticles. The collision of the solitons with different numbers of trapped particles is studied, and it is shown that the collision of the solitons can result in releasing or redistributing the nanoparticles between the solitons. It is demonstrated that the particle mediated interaction between the solitons can affect their dynamics significantly. The reported effects can be successfully used for optical manipulation of nanoparticles and pave a way for efficient control over their dynamics on nanoscale.}


\begin{document}

\maketitle

\section{INTRODUCTION}

A soliton is a nonlinear wave phenomena, which was found in many different systems including hydrodynamics \cite{Water1, Water2}, plasma physics \cite{Plasma}, biology \cite{Biology} and nonlinear optics \cite{Optics1, Optics2}. It was found that solitons existing in many nonlinear optical systems are able to steadily propagate over very long distances \cite{Mollenauer1}. Such properties of the solitons make them very perspective from the practical point of view, see, e.g., \cite{Mollenauer2,Kochetov}.

One of the recent proposed applications of the optical solitons is optical trapping \cite{jetp}. Nowadays optical trapping \cite{Ashkin1, Ivinskaya1, Kostina} and transporting \cite{Ruffner, Petrov, Shalin, Sukhov, Ivinskaya2} is actively developing field and many new effects, that can be used for the trapping, for example such as optical hook \cite{Yue}, is presented. In \cite{jetp} it is proposed to use optical solitons for manipulation of nanoparticles placed in or on the top of the resonator excited by a powerful holding beam. The solution in the form of a bound state of a soliton and a particle is found, and the stability of the states is studied. It is shown that the bound states can be dynamically stable and, thus, can be observed experimentally. 

However, by placing several nanoparticles in the field of a soliton it can be possible to obtain complex soliton-particles states. This can be of interest, because then the solitons can be used as a seeding grains for the formation of clusters of particles. At the same time growing of the particles' number increases the number of degrees of freedom and so can affect the stability of the states strongly. Another important issue is how the trapped particles affect the interaction between the optical solitons. This can be of interest in the light of the use of the multi soliton fields for the control over the particles motion. 

In the present letter we consider the dynamics of the solitons carrying more than one particle and investigate mutual interaction of the solitons with the trapped particles. It is found that the interaction of the solitons with captured particles can lead to different scenarios from a soliton collapse to soliton oscillations depending on the parameters of particles.
The paper is organized as follows. In the next sections we study the interaction of the solitons with trapped particles and show that this interaction can lead to the formation of stable solitons with more than one trapped particle. In the third section we consider the system consisting of two nonlinear resonators interacting only through the particles. It is shown that the particle-mediated interaction can affect the dynamics of the solitons significantly. The main results of the paper are briefly summarized in the conclusion.

\section{COLLISIONS OF SOLITONS WITH CAPTURED PARTICLES}

In this section we consider a nonlinear Fabry-Perot resonator pumped by the coherent light with a dielectric particle, located in the surface. The existence of bistability and solitons in such resonators was shown in \cite{Szoke, Rosanov}. A particle is pulled inside the higher intensity region due to gradient force \cite{Ashkin} and, thus, attracted to bright solitons. In its turn the particle partially screens the holding beam and consequently affects the soliton. The influence of the particle can be strong enough to destroy the soliton. In \cite{jetp} the formation of the soliton-particle bound states is studied and it is shown that in such system stable trapping of particles by the dissipative solitons is possible. However, the collision of the bound states and the formation of the solitons with several trapped particles has never been investigated. This problem is discussed below.

We start with the formulation of the mathematical model describing mutual dynamics of the optical field in a nonlinear resonator and particles. The systems of this type can be described by a generalized nonlinear Schrodinger equation for the optical field. The viscous dynamics of the particle can be obtained via the solution of an ordinary differential equation for the centre mass of the particle \cite{jetp}: 
\begin{eqnarray}
    \label{eq:schrod}
    \dfrac{\partial}{\partial t} E - i C \dfrac{\partial^2}{\partial x^2} E + (\gamma - i \delta + i \dfrac{\alpha}{1 + |E|^2})E = \nonumber \\
    = (1 - \sum\limits_{m} f e^{-(x - \epsilon_m)^2/\omega^2})P, \\
    \dfrac{\partial}{\partial t} \epsilon_m = \eta \dfrac{\partial}{\partial x}  \alpha |E(\epsilon_m)|^2,
    \label{eq:part}
\end{eqnarray}
where $E$ - is a complex amplitude of optical field in the resonator, $C$ - diffraction coefficient, $P$ - complex amplitude of laser pumping, $\gamma$ - decay rate, $\alpha$ - coefficient of nonlinearity; $\delta$ - laser detuning from resonant frequency, $\epsilon$ - coordinate of nanoparticle. Parameter $\omega$ is width of a particle shadow, $f$ is a transparency coefficient of a particle: if $f = 0$, then particle is transparent and if $f=1$ then the particle is opaque. The coefficient $\eta$ defines the ratio of the dragging force acting on the particle to the  field intensity gradient at the point of particle location. Let us note that for a mathematical convenience we use dimensionless variables.

We performed numerical simulations taking two pulses as initial conditions. In a wide range of parameters stable dissipative solitons forms. In the presence of a phase gradient of the holding beam the solitons start moving in the direction of the phase gradient and if the holding beam is $P = P_0 e^{ikx^2}$ then the solitons move towards each other. At some point they collide with particles and the particles stick to them, see \cite{jetp}. Then solitons carrying the particles continue to move towards each other and finally collide at $x \approx 0$. However, the collision can be different depending on the parameters of the solitons.  

\begin{figure}[tbp]
\centering
\fbox{\includegraphics[width=\linewidth]{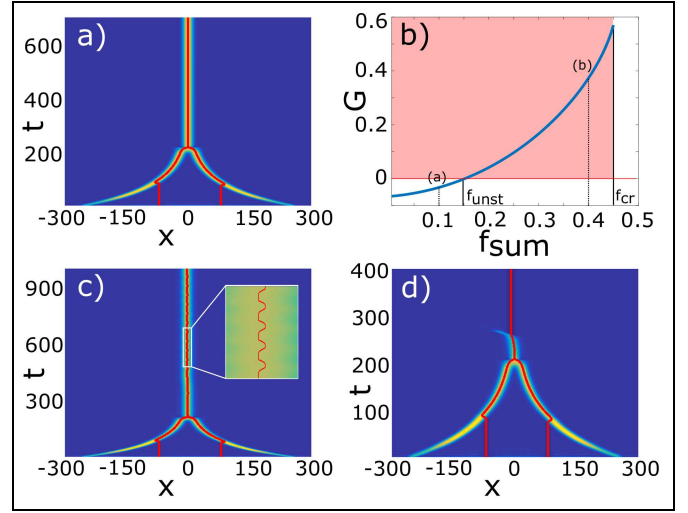}}
\caption{Fig.1 Collision of solitons with trapped particles under inhomogeneous pumping $P = P_0 e^{-i k x^2}$. Parameters: $P_0 = 1$, $k = 0.0002$, $\nu = 0.7$, $\delta = 0.3$, $\gamma = 0.2$, $C = 16$, $\alpha = 3$; (a) As a result of the collision of two initial solitons with trapped particles one soliton is formed, which successfully captures both particles, $f = 0.05$. Corresponding stationary state is stable, see panel (b); (b) The dependence of the instability increment of the resulting soliton with captured particles on the collective transparency of the particles, where $f_{sum} = 2f$, because particles are the same. (c) With more opaque particles ($f = 0.2$) the oscillating soliton with particles is formed as a result of two solitons collision; The inset shows the position of the particles with respect to the soliton. The soliton oscillates because corresponding stationary state is unstable, see panel (b); (d) As a result of the collision both solitons annihilate, $f = 0.25$, because there is no corresponding stationary state.}
\label{fig1}
\end{figure}

One of the possible outcomes is the formation of a single dissipative soliton with two trapped particles. Let us mention here that in our model we disregard direct interaction between the particles. This interaction can be important and affects both the symmetry and the stability of the states but the study of this problem is out of the scope of the present paper. 

Let us note that the aforementioned state can be dynamically stable or unstable depending on the parameters of screening caused by the particles to the holding beam. If both particles are transparent enough, and the total losses of pumping intensity introduced by particles are relatively weak ($f_{sum} < f_{unst}$), the resulting soliton is able to capture both particles placed in the center of system by initial solitons, see fig.\ref{fig1}(a). The corresponding stationary state of the resting soliton with captured particles is dynamically stable, see fig.\ref{fig1}(b). 

The situation changes if total losses overcome the critical value $f_{sum} > f_{unst}$. In this case the shadow created by particles repels the soliton from the particles. At the same time the particles continue to get attracted to the soliton. Moreover, the phase gradient of the holding beam pushes the soliton to the equilibrium point $x=0$. The interplay of these factors destabilizes the soliton with two particle. Instead of this an oscillating state forms, see fig.\ref{fig1}(c); where on the inset it is seen that the trapped particles oscillate around the equilibrium point.

If the shadow created by particles is even more significant $f_{sum} > f_{cr}$, then resulting soliton collapses releasing the particles, see fig.\ref{fig1}(d). In this case the total losses of the pumping intensity introduced by the particles are too large, and pumping can't support the soliton. The corresponding stationary state of the resting soliton with particles does not exist, see fig.\ref{fig1}(b). In other words, both initial solitons annihilate after the collision.

\section{INTERACTION OF SOLITONS THROUGH RESCATTERING ON PARTICLES}

Now we consider two nonlinear wide-aperture resonators separated by a relatively thin gap. Each of the resonators is pumped by a holding beam, and we assume that the modes in the resonators are well localized, so that the resonators do not interact with each other directly. However, if a particle is placed between them, then it feels the evanescent fields of both the resonators modes. That is why the particle can get attracted to the maximal intensity region of the field in each of the cavities. At the same time coupling to the particles decreases coupling of the holding beam to the guided mode of the cavity. Thus, the modes of the cavities can interact through the particles placed between them.

Here we use the model analogous to one from (\ref{eq:schrod}-\ref{eq:part}) describing the interaction of a particle with optical field. The fields in the resonators do not interact with each other directly, so the equations for $E_1$ are independent to $E_2$ and vise versa, but both fields do interact with the particles placed between resonators. The self-consistent system of equations takes the form:

\begin{eqnarray}
    \dfrac{\partial}{\partial t} E_1 - i C \dfrac{\partial^2}{\partial x^2} E_1 + (\gamma - i \delta - i \dfrac{\alpha}{1 + |E_1|^2})E_1 = \nonumber \\
    = (1 - \sum\limits_{m} f e^{-(x - \epsilon_m)^2/\omega^2})P_1, \\
    \label{eq:schrod1}
    \dfrac{\partial}{\partial t} E_2 - i C \dfrac{\partial^2}{\partial x^2} E_2 + (\gamma - i \delta - i \dfrac{\alpha}{1 + |E_2|^2})E_2 = \nonumber \\
    = (1 - \sum\limits_{m} f e^{-(x - \epsilon_m)^2/\omega^2})P_2, \\
    \label{eq:schrod2}
    \dfrac{\partial}{\partial t} \epsilon_m = \eta \dfrac{\partial}{\partial x}  \alpha |E(\epsilon_m)|^2,
    \label{eq:part1}
\end{eqnarray}
where $E_1$ and $E_2$ - are complex amplitudes of optical fields in the resonators, other parameters are the same as in equations (\ref{eq:schrod}-\ref{eq:part}).

\begin{figure}[t]
\centering
\fbox{\includegraphics[width=\linewidth]{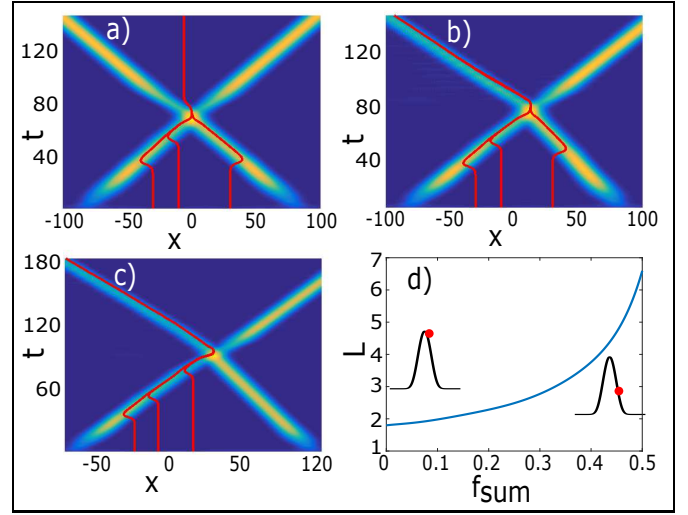}}
\caption{Fig. 2. Interaction of solitons through rescattering on the particles. Parameters: $P_1 = P_0 e^{i k_1 x}$,$P_2 = P_0 e^{-i k_2 x}$, $f = 0.15$, $\nu = 0.7$, $P_0 = 1$, $\delta = 0.3$, $\gamma = 0.2$, $C = 16$, $\alpha = 3$. (a) In a result of collision of solitons moving with the similar velocities and carrying trapped particles all particles get released. $k_1 = k_2 = 0.4$; (b), (c) Initial solitons with trapped particles move with different velocities ($k_1 = 0.4$, $k_2 = 0.3$) and in a result of collision all particles get captured by the slower soliton; (d) Dependence of the distance between the particles and the soliton center on the $f_{sum}$; $k = 0.04$, other parameters are the same as in (a).}
\label{fig2}
\end{figure}

Interaction of solitons through the rescattering on the particles can lead to different scenarios of the further dynamics after the collision of solitons with trapped particles. By <<collision>> we mean the solitons to overlap, but not interact directly. However, the overlap of the solitons means that both of the solitons interact with the same particle, and this can cause an effective interaction between them.

We studied numerically the considered system of equations with three particles. The initial conditions are taken as two pulses separated by a relatively large distance. The holding beams are taken in the form $P_1 = P_0 e^{i k_1 x}$, $P_2 = P_0 e^{-i k_2 x}$ causing the solitons to move towards each other, see fig.\ref{fig2}. Before the overlap the solitons collide with the particles and trap them. The parameters of the solitons are chosen  to provide stable trapping.

As an example, we consider the left soliton capturing two particles and the right soliton capturing only one particle. If $k_1 = k_2$, then both the solitons move towards each other with similar velocities (the velocities are the same if the amount of particles captured by the solitons is the same). We found that one of the possible interaction scenarios is the realase of the particles by the solitons after their overlap, see fig.\ref{fig2}(a). Such behavior can be explained by the fact that because of the collision of the solitons with trapped particles the intensity of the pumping field decreases, and the solitons are not able to capture particles any more. After releasing the particles both solitons are restored. This effect can be used to collect the particles over the trajectories of the solitons and then release all of them in one point, where the solitons meet each other.

\begin{figure}[t]
\centering
\fbox{\includegraphics[width=\linewidth]{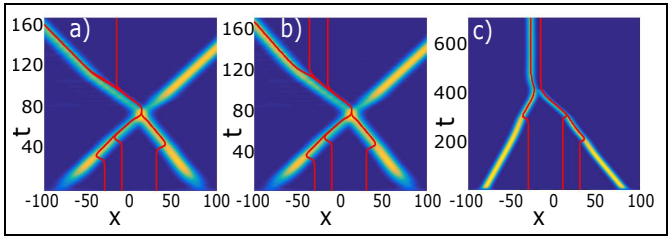}}
\caption{Fig.3. Interaction of solitons through rescattering on the particles. (a) As a result of the collision the slower soliton captures all the particles, but it is able to hold only two particles, and one particle gets released. Parameters: $k_1 = 0.4$, $k_2 = 0.3$, $f = 0.156$, $\nu = 0.7$, $P_0 = 1$, $\delta = 0.3$, $\gamma = 0.2$, $C = 16$, $\alpha = 3$; (b) The slower soliton captures all the particles, but it is able to hold only one particle, that is why two particles get released after the collision. $f = 0.1564$ and other parameters are the same as in panel (a); (c) The faster soliton collapses releasing the particles and the slower soliton gets stopped by the released particles. Parameters: $k_1 = 0.2$, $k_2 = 0.3$, $f = 0.15$, $\nu = 0.1$, $C = 4$.}
\label{fig3}
\end{figure}

If $k_1 > k_2$, then the solitons move towards each other with different velocities. We found out that after a collision the slower soliton takes away all particles from the faster soliton, see fig.\ref{fig2}(b). If the slower soliton is able to hold steadily all the particles, then the soliton carries them further. This result can be explain by the fact that the faster soliton needs larger force to capture a particle, then the slower one. It should be mentioned that after the collision the slower soliton always captures all the particles, even if the initial slower soliton does not have trapped particles at all, see fig.\ref{fig2}(c). 

It is interesting to note, that the more particles soliton carries, the bigger distance between the center of the soliton and the particles. We found numerically dependence of that distance ($L$) on the collective transparency of the trapped particles, see fig.\ref{fig2}(d). It  can be seen that even almost transparent particles (with $f \approx 0$) are shifted from the center of the soliton, and more opaque particles are shifted even more from the center to the tale of the soliton.

We also study the case, when the slower soliton captures so many particles that the intensity of the solitons decreases significantly, and the soliton cannot carry all the particles, see fig.\ref{fig3}(a-b). Then one of the particles gets released, this increases the soliton intensity, and the rest of the particles stay trapped in the soliton, fig.\ref{fig3}(a). It can happen that releasing only one particle is not enough, and more particles decouple from the soliton; in fig.\ref{fig3}(b) the case of two particles leaving the soliton is shown. 

One more scenario of the soliton collision is the collapse of the faster soliton and the release of the particles trapped in it. At the same time the intensity of the slower soliton is not sufficient to capture all the particles. Therefore, as it is shown in fig.\ref{fig3}(c), one of the particles is trapped by a slow soliton, and two particles are released at the point of the death of the faster soliton. In the case shown in fig.\ref{fig3}(c) the phase gradient of the holding beam pushes the soliton towards the released particles, but the losses introduced by the particles push the soliton in the opposite direction. As a result, a nearly motionless state of the soliton and the particles forms in the system.

It should be noted here that the direct interaction between the solitons can make the dynamics much richer, but we decided to consider a simpler case to emphasise the role of the particles-mediated interaction in the behaviour of the solitons. The more complex problem accounting for the direct inter-soliton interaction has to be considered separately.

\section{CONCLUSION}
In conclusion we would like to summarize the main results of the paper. The direct interaction of dissipative solitons in the pumped nonlinear Fabry-Perot resonator with trapped particles is considered. It is shown that in a result of two soliton collision with trapped particles three scenarios are possible: first, solitons collapse with releasing the particles; second, oscillating regime of particles trapping by the solitons; and third, successful capturing of particles by the soliton formed after the soliton collision. The dependence of all possible cases on particle parameters is considered.

The indirect interactions of solitons are studied. The interaction mediated by the particles trapped on the solitons is considered. It is shown that trapping and the result of the soliton collision depend on the velocities of the solitons.The particles are trapped by the slower soliton provided that the intensity of the solitons are the same. The case of the solitons of different intensities is not discussed in the main text but here we should mention that, of course, the amplitude of the solitons affects trapping strongly and if the faster soliton has significantly higher amplitude then it traps particles more efficiently.

The main results of the paper are that the solitons and the particles can form stable bound states and that the solitons can be used for a flexible control over the particles. For example, it is possible to collect all the particles in the system by two counter-propagating solitons and then release the particles at the point of the soliton collision. Since the position of the soliton  collision can be controlled by the solitons, it opens a possibility to create a cluster of particles in a desirable point in the system. Alternatively, it is possible to trap all the particles to a soliton to bring them to the left or to the right edge of the cavity. Finally, it is shown that the interaction between the solitons and the particles can result in the formation of bound states moving with the velocity much lower than the velocity of the solitons without trapped particles. Such accurate many particle manipulation can be used, for example, in manufacturing of high efficient antireflective coatings for solar sells \cite{Voroshilov}.

The work was supported by the Russian Foundation for Basic Research (Projects No. 18-02-00414, 18-52-00005). The calculations of the soliton dynamics are partially supported by the Russian
Science Foundation (Project No. 18-72-10127).


\begin{thebibliography}{99}

\bibitem{Water1}
D.\,H. Peregrine, The ANZIAM Jour. {\bf 25}, Issue 1 (1983).

\bibitem{Water2}
R. Hirota and J. Satsuma, J. Phys. Soc. Japan. {\bf 40}, 2 (1976).

\bibitem{Plasma}
N.\,J. Zabusky and M.\,D. Kruskal, Phys. Rev. Lett., {\bf 15}, 6 (1965).

\bibitem{Biology}
H. Kuwayama and S. Ishida, Sci. Rep. {\bf 3}, 2272 (2013).

\bibitem{Optics1}
J.\,T. Taylor, Ed., Optical Solitons - Theory and Experiment (Cambridge University Press,
New York, 1992).

\bibitem{Optics2}
Y.\, S.\, Kivshar and G. Agrawal, Optical Solitons - From Fibers to Photonic Crystals (Academic Press, 2003).

\bibitem{Mollenauer1}
L.\,F. Mollenauer and K. Smith, Opt. Lett., {\bf 13}, 8 (1988).

\bibitem{Mollenauer2}
L.\,F. Mollenauer, E. Lichtman, M.\,J. Neubelt and G.\,T. Harvey, Conference on Optical Fiber Communication, {\bf 4}, PD8 (1993).

\bibitem{Kochetov}
B. Kochetov, I. Vasylieva, A. Butrym and V.\,R. Tuz, Phys. Rev. E, {\bf 99}, 052214 (2019).

\bibitem{jetp}
D.\,A. Dolinina, A.\,S. Shalin, A.\,V, Yulin, arXiv:1909.00618[physics.optics].

\bibitem{Ashkin1}
A. Ashkin, Phys. Rev. Lett. {\bf 24}, 156
(1970).

\bibitem{Ivinskaya1}
A. Ivinskaya, M.\,I. Petrov, A.\,A. Bogdanov, I. Shishkin, P. Ginzburg and A.\,S. Shalin, Light: Science and Applications {\bf 6}, e16258 (2017).

\bibitem{Kostina}
N. Kostina, M. Petrov, A. Ivinskaya, S. Sukhov, A. Bogdanov, I. Toftul, M. Nieto-Vesperinas, P. Ginzburg and A. Shalin, Phys. Rev. B {\bf 99}, 125416 (2019).

\bibitem{Ruffner}
D.\,B. Ruffner and D.\,G. Grier, Phys.~Rev.~Lett. {\bf 109}, 163903 (2012).

\bibitem{Petrov}
M.\,I. Petrov, S.\,V. Sukhov, A.\,A. Bogdanov, A.\,S. Shalin and A. Dogariu, Laser Photonics Rev. {\bf 10}, 116 (2016).

\bibitem{Shalin}
A.\,S. Shalin and S.\,V. Sukhov, Plasmonics, {\bf 8}, 2 (2013).

\bibitem{Sukhov}
S. Sukhov, A. Shalin, D. Haefner and A. Dogariu, Opt. Exp., {\bf 23}, 247 (2015).

\bibitem{Ivinskaya2}
A. Ivinskaya, N. Kostina, A. Proskurin, M.\,I. Petrov, A.\,A. Bogdanov, S. Sukhov, A.\,V. Krasavin, A. Karabchevsky, A.\,S. Shalin and P. Ginzburg, ACS Photonics, {\bf 5}, 4371 (2018).

\bibitem{Yue}
L. Yue, O.\,V. Minin, Z. Wang, J.\,N. Monks, A.\,S. Shalin, and I.\,V. Minin, Opt.~Lett. {\bf 43}, 4 (2018).

\bibitem{Szoke}
A. Szöke, V. Daneu, J. Goldhar, and N.\,A. Kurnit, Appl.~Phys.~Lett. {\bf 15}, 376 (1969).

\bibitem{Rosanov}
N.\,N. Rosanov and G.\,V. Khodova, Opt.~Soc.~Am.B {\bf 7}, 6 (1990).

\bibitem{Ashkin}
A. Ashkin, J.\,M. Dziedzic, J.\,E. Bjorkholm and Steven Chu, Opt.~Lett. {\bf 11}, 5
(1986)

\bibitem{Voroshilov}
P.\,M. Voroshilov, C.\,R. Simovski, P.\,A. Belov, and A.\,S. Shalin, J.~Appl.~Phys, {\bf 117}, 203101 (2015).

\end{thebibliography}
\end{document}